\documentclass[preprint,12pt]{elsarticle}

\usepackage{amssymb}
\usepackage{amsmath}
\usepackage{graphicx}
\usepackage{amsfonts}

\def\H{\mathsf{H}}
\def\K{{\cal K}}
\def\hP{\hat\Pi}

\journal{Physics Letter B}

\begin{document}

\begin{frontmatter}


\title{Wheeler--De Witt equation and AdS/CFT correspondence}

\author{Francesco Cianfrani, Jerzy Kowalski-Glikman}
\address{Institute
for Theoretical Physics, University of Wroc\l{}aw, Pl.\ Maksa Borna
9, Pl--50-204 Wroc\l{}aw, Poland}
\date{\today}


\begin{abstract}
The radial Wheeler--De Witt equation on the asymptotically AdS
spacetime proposed in \cite{Freidel:2008sh} has as its semiclassical
solution the wave function that asymptotically satisfies the
conformal Ward identity, exemplifying the AdS/CFT correspondence. In
this paper we show that this results holds also in the case of a
complete quantum solution of the radial Wheeler--De Witt equation.
It turns out that if the wavefunction is expanded in the parameter
$\rho$ with $\rho\rightarrow0$ defines the asymptotic boundary of the
spacetime, the quantum loop corrections to the semiclassical wave
are of subleading order.
\end{abstract}


\end{frontmatter}



\section*{Introduction}

The AdS/CFT correspondence is a deep and beautiful relation between
quantum gravity theory in the bulk of an asymptotically Anti de
Sitter space and conformal field theory on the boundary of this
space. It has been first conjectured in the context of string theory
\cite{Maldacena:1997re} on AdS$_5$ $\times$ S$^5$ background, but it
has soon been realized that it might be a generic property of any
quantum gravity theory. This interpretation of AdS/CFT is implicit
in the seminal Witten's paper \cite{Witten:1998qj} (see also
\cite{Gubser:1998bc}) and has been further supported by
semiclassical investigations reported in the series of papers
\cite{Henningson:1998gx}, \cite{de Boer:1999xf},
\cite{deBoer:2000cz}, \cite{de Haro:2000xn},
\cite{Papadimitriou:2004ap}.

One of the most interesting aspects of these investigations was the
discovery of a direct relation between the radial flow off the
boundary in the bulk with the renormalization group flow in the
boundary theory \cite{de Boer:1999xf}. This relation was established
by using the Hamilton-Jacobi theory that makes it possible to cast
the flow equation of (super) gravity into the form of the
Callan-Symanzik equations. It was already suggested in \cite{de
Boer:1999xf} that a natural generalization of its results would be
to consider the full Wheeler-De Witt equation, especially in view of
the formal similarity between the Wheeler-De Witt equation and the
right hand side of the Polchinski's exact renormalization group
equation \cite{Polchinski:1983gv}.

This research program has started with an interesting paper
\cite{Freidel:2008sh}, in which it was shown that AdS/CFT
correspondence can be established in the case of the ``wave function
of the universe'', a solution of the radial Wheeler-De Witt equation
\cite{DeWitt:1967yk}, and that such solutions provide a simple and
systematic way to derive the local action in the asymptotic
expansion. As shown in that paper in the case of an asymptotically
AdS space the radial Wheeler--De Witt equation is formally exactly
the same as the standard one (see e.g., \cite{Isham:1995wr} for
review). The difference in the physical interpretation of these two
equations is that while the standard Wheeler--De Witt equation is
related to the Hamiltonian constraint of general relativity, related
with evolution in time, the radial equation describes a constraint
associated with radial expansion of a ``constant space-time radius
surface.'' Thus, by solving this latter equation, and taking the
limit of infinite radius one can investigate the property of
asymptotic wave function. It has been argued in
\cite{Freidel:2008sh} that in such a limit the wave function indeed
possesses desired properties, satisfying the correct conformal Ward
identity.

In particular, the correspondence reads
\begin{equation}\label{Fcorr}
\Psi_\rho\left(\frac{\gamma}{\rho^2}\right)=e^{\frac{i}{\kappa}S^{(d)}\left(\frac{\gamma}{\rho^2}\right)}Z_++e^{-\frac{i}{\kappa}S^{(d)}\left(\frac{\gamma}{\rho^2}\right)}Z_-,
\end{equation}
$\Psi_\rho(\gamma)$ being the wave function on the $D$-dimensional
spatial slice $\Sigma_\rho$ having metric $\gamma_{ab}$,  while the
functionals $Z_\pm$ are solutions of the (nonanomalous) Ward
identities associated with conformal invariance on $\Sigma_\rho$.
The local action $S^{(d)}(\gamma)$ can be expanded in the limit of
small AdS radius $\rho$  and for $\rho\rightarrow 0$ has the
following properties (at least for $d<5$):
\begin{itemize}
\item the terms with negative powers of $\rho$ coincide with minus the counterterms $S_{ct}(\gamma)$ adopted in holographic renormalization to regularize the gravitational action \cite{de Haro:2000xn}\footnote{
This result found an explanation in \cite{Papadimitriou:2010as}, where the correspondence between the terms responsible for the finiteness of the action and the Hamilton-Jacobi formulation has been elucidated.
};
\item the term of the $\rho^0$ order gives the anomaly $A_{d}$ of the conformal theory on $\Sigma_\rho$.
\end{itemize}
Therefore, one can write
\begin{equation}
e^{\frac{i}{\kappa}S^{(d)}\left(\frac{\gamma}{\rho^2}\right)}Z=e^{-\frac{i}{\kappa}S_{ct}\left(\frac{\gamma}{\rho^2}\right)}\widetilde{Z},
\end{equation}
in which $\widetilde{Z}$ satisfies the conformal anomalous Ward identity.

One of the new features of \cite{Freidel:2008sh} is that the
correspondence holds between a quantum gravity model and two
conformal theories. This is due to the fact that the Wheeler-De Witt
equation is of second order and a solution is specified by two
coefficients $Z_\pm$. One finds out the standard AdS/CFT
correspondence by fixing $Z_-=0$ and by performing a semiclassical
limit for $\Psi_\rho$, {\it i.e.}
\begin{equation}
\Psi_\rho(\gamma)=\int_{g|_{\Sigma_\rho}=\gamma} Dg\; e^{\frac{i}{\kappa}S(g)}\sim e^{\frac{i}{\kappa}S_{\rho}(\gamma)},
\end{equation}
$S_{\rho}(\gamma)$ being the gravitational action evaluated at the classical solution having as initial condition the metric $\gamma$ on $\Sigma_\rho$. This way, one can rewrite (\ref{Fcorr}) as follows
\begin{equation}
e^{\frac{i}{\kappa}S_{\rho}\left(\frac\gamma{\rho^2}\right)}=e^{-\frac{i}{\kappa}S_{ct}\left(\frac{\gamma}{\rho^2}\right)}\widetilde{Z}_+,
\end{equation}
and by moving  to the left-hand side the factor
$e^{-\frac{i}{\kappa}S_{ct}}$ one gets the standard correspondence
between the regularized gravitational action and the solution of the
Ward identity for a (possibly anomalous) conformal theory.

It is should be stressed that the approach presented in
\cite{Freidel:2008sh} automatically accounts for both holographic
renormalization \cite{Henningson:1998gx}, \cite{de Boer:1999xf},
\cite{deBoer:2000cz}, \cite{de Haro:2000xn},
\cite{Papadimitriou:2004ap} and the anomaly of the conformal theory,
which, in the standard AdS/CFT paradigm, are unrelated.

The argument of \cite{Freidel:2008sh}  is however not complete since
renormalization effect are not taken into account there. Such
effects necessarily arise because the Wheeler--De Witt operator
contains two functional derivatives acting at the same spacetime
point, whose action produce terms proportional to the $n$th
derivative delta function at zero, $\delta^{(n)}(0)$, which should
be carefully renormalized \cite{Mansfield:1994}. It is not clear
therefore if the results of \cite{Freidel:2008sh} stands when the
presence of such terms is taken into account. This is the question
we would like to address in this paper.

Some time ago in the paper \cite{KowalskiGlikman:1996ad} we showed
that there is a way to regularize Wheeler--De Witt equation and to
renormalize its action on wave functions.  The method of that paper
can be readily applied in the present context. But let us start
 reviewing briefly the argument presented in \cite{Freidel:2008sh} showing the formal
coincidence between the radial Wheeler--De Witt equation and the
standard one.

In a $D+1$-dimensional space-time manifold $\mathcal{M}$ endowed
with metric $g_{\mu\nu}$ and with  boundary $\Sigma$ having the
induced metric $\gamma_{ab}$, the action for gravity which leads to
the  well-posed variational principle reads
\begin{equation}\label{action}
S(g)=-\frac{1}{2\kappa}\int_{\mathcal{M}}d^{D+1}x\sqrt{g}(R(g)-2\Lambda)+\frac{\epsilon}{\kappa}\int_\Sigma d^Dx\sqrt{\gamma}K,
\end{equation}
where $R$ denotes the scalar curvature, and $K=\gamma^{ab}K_{ab}$ is
the trace of the extrinsic curvature on $\Sigma$. The cosmological
constant $\Lambda$ is related to the cosmological length scale
$\ell$ as follows
\begin{equation}
\Lambda=-\epsilon\frac{D(D-1)}{2\ell^2}
\end{equation}
$\epsilon$ being $-1$ ($+1$) for $\Sigma$ spacelike (timelike) and
$\kappa=8\pi G$.

The on-shell variation of the action (\ref{action}) under a bulk
diffeomorphism reads
\begin{equation}
\delta_\xi S(g)=-\frac{1}{2\kappa}\int_\Sigma d^Dx \sqrt\gamma\xi \left[R(\gamma)-2\Lambda+\epsilon(K^2-K^a_bK_a^b)\right]
\end{equation}
$\xi$ being the infinitesimal parameter of the transformation.  The
expression inside square brackets in the equation above coincides
formally with the superhamiltonian, the only difference being the
presence of the factor $\epsilon$ (which in ADM formalism is fixed
to $-1$ corresponding to the spacelike boundary). Hence, the Ward
identity associated with bulk diffeomorphisms invariance tell us
that the associated operator $\H$ vanishes when acting on the wave
function $\Psi(\gamma)=\int_{g|_{\Sigma}=\gamma} Dg\;
e^{\frac{i}{\kappa}S(g)}$. Thus, the radial evolution is dictated by
an equation which is essentially identical to the Wheeler-De Witt
one.

Following  \cite{Freidel:2008sh} we therefore write the radial
Wheeler--De Witt operator $\H$ as follows
\begin{equation}\label{1}
   \H = -\kappa^2\hP\circ\hP(x)+ R[\gamma](x) + \frac{D(D-1)}{\ell^2},
\end{equation}
with
\begin{equation}
\hP\circ\hP(x)=\hP^a_b(x)\hP^b_a(x) - \frac{\hP(x)\hP(x)}{D-1}.
\end{equation}
In this equation $\hP$ is the momentum operator
\begin{equation}\label{2}
   \hP^a_b(x) \equiv \frac{2}{\sqrt{\gamma(x)}}\, \gamma_{bc}\, \frac{\delta}{\delta\gamma_{ac}(x)}, \quad \hP \equiv
   \hP^a_a\,,
\end{equation}
$\gamma_{ab}$ is the metric on $D$ dimensional space $\Sigma_\rho$
of constant ``radius'' $\rho$. Notice that in writing down (\ref{1})  we
made use of a particular ordering defined by (\ref{2}), resolving
therefore the ordering ambiguity. This particular ordering has the
virtue that, as we will see below, when $\H$ acts on the volume
${\cal V} = \int_{\Sigma_\rho} \sqrt{\gamma} \, d^Dx$ the result is finite and
does not require  renormalization.

As it stands, eq.\ (\ref{1}) is meaningless because it involves the
product of two functional derivatives defined at the same point, and
clearly requires regularization. Following the idea of
\cite{KowalskiGlikman:1996ad} we define the regularized Wheeler--De
Witt operator $\H^{reg}$ with the help of heat kernel, to wit
\begin{equation}\label{3}
\H^{reg} =-\kappa^2 \hP\circ_{\K}\hP(x)+ R[\gamma](x) + \frac{D(D-1)}{\ell^2}
\end{equation}
with
$$\hP\circ_{\K}\hP(x)=\int_{\Sigma_\rho} d^Dy \K(x,y;t)\sqrt{\gamma(x)}\left(\hP^a_b(x)\hP^b_a(y) -
\frac{\hP(x)\hP(y)}{D-1}\right)$$

The heat kernel satisfies the equation
\begin{equation}\label{4}
    \frac{\partial}{\partial t}\K(x,y;t) = \nabla^2_{(x)} \K(x,y;t) + \xi
R(x) \K(x,y;t)
\end{equation}
with the  initial condition
\begin{equation}\label{5}
    \lim_{t \rightarrow 0} \K(x,y;t) = \frac{\delta^{(3)}(x-y)}{\sqrt{\gamma(x)}}
\end{equation}
Equation (\ref{4}) can be solved perturbatively in powers of $t$. To
the order which will be of  interest in the present context, the
solution reads \cite{Parker:1984dj} (see \cite{Codello:2013} for a
recent discussion)
$$K(x,y;t)  = \frac{\exp\left(\left(-\frac1{4t}\gamma_{ab}(x) -
\frac1{24}R_{ab}(x)\right)\Delta^a\Delta^b\right)}{(4\pi t)^{\frac
D2}}\bigg\{ 1 + t\left(\xi - \frac16\right) R(x) + t^2$$
\begin{equation}
\left[{\frac16}\left(\xi -\frac15\right) \nabla^2 R(x)
+{\frac 12}\left(\xi -\frac16\right)^2 R^2(x)+
\frac1{60}R_{ab}(x)R^{ab}(x) -\frac1{180} R^2(x)\right] +
O(t^3)\bigg\}\label{6}
\end{equation}
with $\Delta^a=x^a-y^a$.

 Then, the coincidence limit $t\rightarrow 0$ is performed, such that
 all the terms with positive powers of $t$ are avoided. However, some infinities
 generically arise because of the presence of negative powers of $t$. These terms
 can be treated according with the procedure defined in \cite{Mansfield:1994}
 via analytic continuation and they define renormalized (dimensionful) constant.

Having established the form of Wheeler--De Witt radial operator we
define the wave function $\Psi(\gamma)$ on a radial slice
parametrized by radius $\rho$ to be a solution of Wheeler--De Witt
equation
\begin{equation}\label{7}
    \H \Psi =0
\end{equation}
It should be stressed that $\Psi$ encodes all the information  on
quantum space-time and in particular, as argued in
\cite{Freidel:2008sh} its asymptotic properties can be read off from
the infinitely rescaled wave function $\Psi(\rho^{-2}\gamma)$ in the
limit when $\rho$ is going to zero.

To investigate this asymptotic behavior of the solutions of
Wheeler--De Witt equation it is convenient to rescale variables
\begin{equation}\label{8}
   \gamma_{ab} \rightarrow \rho^{-2}\gamma_{ab}
\end{equation}
so that the radius $\rho$ is explicit in the Wheeler--De Witt
operator:
\begin{equation}\label{9}
  \H^{reg}_\rho = -\kappa^2 \rho^{D}\hP\circ_{\K_\rho}\hP(x) + \rho^2\, R[\gamma](x) +
\frac{D(D-1)}{\ell^2}
\end{equation}
with the rescaled wave function $\Psi(\rho^{-2}\gamma)\equiv\Psi_\rho(\gamma)$ being a solution of
\begin{equation}\label{10}
   \H_\rho\, \Psi_\rho =0
\end{equation}
The crucial difference between (\ref{8}) and the analogous formula
in \cite{Freidel:2008sh} is  that now the heat kernel acquires the
dependence on the radius $\rho$. Let us investigate if this would
lead to any modifications of the results presented there.

It is quite remarkable that the rescaled heat kernel has almost the
same form as the original one.  In fact, one can rescale the heat
kernel time $t \rightarrow t'=t\rho^2$ to find
\begin{equation}\label{11}
   \K_\rho(x,y;t) = \rho^D\, \K(x,y,t')
\end{equation}
so that (\ref{9}) becomes
\begin{equation}\label{12}
 \H^{reg}_\rho = -\kappa^2 \rho^{2D}\hP\circ_{\K}\hP(x) + \rho^2\, R[\gamma](x) +
\frac{D(D-1)}{\ell^2}\,,
\end{equation}
where the expansion of the kernel is given in (\ref{6}).

The asymptotic behavior of the solution (\ref{10}) can be
investigated by writing
$\Psi_{\rho}=\Psi^{(0)}_\rho\Psi^{(1)}_\rho\Psi^{(2)}_\rho\ldots$
and by performing an expansion of $\H_\rho\, \Psi_\rho$ in powers of
$\rho$.

\section*{D=2}

Let us use the expression (\ref{12}) in the case of the of
AdS$_3$/CFT$_2$. Following \cite{Freidel:2008sh} we use the ansatz
\footnote{It is worth noting that $S_0$ and the analogous terms
appearing at higher orders of the asymptotic expansions are actually
infinite since the integration extends over the non compact space
$\Sigma_\rho$, which for AdS$_D$ is $R\times S_{D-2}$.}
\begin{equation}\label{j1}
\Psi_\rho(\gamma)  = e^{i S^0}\, Z_+(\gamma) + e^{-i S^0}\,
Z_-(\gamma)\,,\quad S^0= \frac{1}{\kappa\ell\rho^2}\, \int_{\Sigma_\rho}
\sqrt{\gamma}
\end{equation}
so that $e^{\pm iS^0}$ satisfies the equation
$$
\kappa^2 \rho^{4} \hP\circ_{\K}\hP(x) \, e^{\pm iS^0}= \frac{2}{\ell^2}\, e^{\pm iS^0}\,.$$
This is easy to see noticing that as a result of the ordering we use
there is no second derivative of $S^0$ term resulting from the left
hand side of this expression. Then $Z_\pm$ satisfy the equation
\begin{equation}\label{j2}
\left\{-\kappa^2 \rho^{2}\,\hP\circ_{\K}\hP(x)\,   \pm 2i\frac\kappa\ell\,
\hP(x)\right\} Z_\pm =-\, R[\gamma](x)\, Z_\pm
\end{equation}
If we neglect the first term for a moment we find the conformal Ward
identity
\begin{equation}\label{j3}
    \hP\, Z_\pm = \pm i\frac{\ell}{2\kappa}\,  Z_\pm\,,
\end{equation}
with central charge $c=12\ell/\kappa=3\ell/2G$.

There are two ways to write the solution of the equation above.
The first one is discussed also in \cite{Freidel:2008sh} and it reads
\begin{equation}\label{25}
Z_\pm(e^{2\phi}\hat\gamma)=e^{\pm i\frac{\ell}{2\kappa}S_{L}(\phi,\hat{\gamma})}Z(\hat{\gamma}),\qquad S_L(\phi,\hat{\gamma})=\int_\Sigma \sqrt{\gamma}(\phi\nabla^2\phi+\phi R[\gamma])d^2x,
\end{equation}
in which $\gamma_{ab}=e^{2\phi}\hat\gamma_{ab}$ and $\det{\hat\gamma}=1$.

Equivalently the solution of Eq.(\ref{j3}) can be written in terms of the non local Polyakov action \cite{Polyakov87}, {\it i.e.}
\begin{equation}
S_{Pol}=\int d^2x d^2x' (\sqrt{\gamma}R)(x)G(x,x')(\sqrt{\gamma}R)(x'),
\end{equation}
in which the propagator $G(x,x')$ is the inverse of the Laplacian operator:
\begin{equation}
\sqrt{\gamma(x)}\nabla^2_x G(x,x')=\delta(x-x').
\end{equation}
The solution reads
\begin{equation}
Z_\pm(\gamma)=e^{\mp i\frac{\ell}{8\kappa}S_{Pol}}.
\end{equation}

It is easy to see that in both cases asymptotically the contribution to the
equation (\ref{j3}) resulting from the first term in (\ref{j2}),
including the loop corrections, is negligible. Indeed $Z_\pm$ does
not depend on $\rho$ and therefore the first term in (\ref{j2}) is
negligible compared to other two in the limit $\rho\rightarrow0$.
Thus there is no correction to $Z_\pm$ in the leading order
resulting from this term, in particular from those that result from
the coincidence limit of the heat kernel, that we will call `loop
corrections' (see below.) Such corrections will be, of course
present, if the terms of higher order in $\rho$ are taken into
account, but unfortunately, due to the complexity of such term we
have not be able to calculate their explicit form.

\section*{D=3}

Let us now turn to the $D=3$ case, which corresponds to the physical
(at least at low energies) four dimensional, asymptotically AdS
spacetime, i.e., to the AdS$_4$/CFT$_3$ correspondence. In this case
we use the ansatz
\begin{equation}\label{j4}
\Psi_\rho(\gamma)  = e^{i (S^0+ S^1)}\, Z_+(\gamma) + e^{-i (S^0+
S^1)}\, Z_-(\gamma)\,,
\end{equation}
where $S^0$ and $S^1$ are the leading and next-to-leading order
solutions of the radial Wheeler--De Witt equation
$$
H_\rho\Psi_\rho(\gamma)  =0\,,
$$
\begin{equation}\label{j5}
H_\rho =-\kappa^2 \rho^{6}\,\hP\circ_{\K}\hP(x) + \rho^2\, R[\gamma](x) +
\frac{6}{\ell^2}\,.
\end{equation}
Similarly to the $D=2$ case we assume that $e^{iS^0}$ satisfies the
equation
$$
\left\{-\kappa^2 \rho^{6}\,\hP\circ_{\K}\hP(x)  + \frac{6}{\ell^2}\right\}e^{iS^0}
=0\,,
$$
and we find that
\begin{equation}\label{j6}
    S^0=\frac{2}{\kappa\ell\rho^3} \int_{\Sigma_\rho} \sqrt{\gamma}\, d^3x\,.
\end{equation}
Substituting this to (\ref{j5}) we get
\begin{equation}\label{j7}
\bigg[\kappa^2 \rho^{4}\hP\circ_{\K}\hP(x) -  R[\gamma](x)\mp
2\frac{i\kappa\rho}{\ell}\, \hP(x) \bigg] e^{\pm i S^1}\, Z_\pm =0
\,.
\end{equation}
The first term in (\ref{j7}) seems again to be negligible in the
limit $\rho\rightarrow0$, so let us use the ansatz
$$
\left(R[\gamma](x)\pm 2\frac{i\kappa\rho}{\ell}\, \hP(x) \right)
e^{\pm i S^1}=0,
$$
from which we get
\begin{equation}\label{j8}
   S^1 =\frac{\ell}{2\kappa\rho}\, \int_{\Sigma_\rho} \sqrt{\gamma}\, R[\gamma]
   d^3y\,.
\end{equation}
Then, still neglecting the first term in (\ref{j7}), we find that
\begin{equation}\label{j9}
    \hP\, Z_{\pm} =0\,,
\end{equation}
which expresses the conformal invariance of $Z_{\pm}$. A solution of
this equation is given be the Chern-Simons action
\cite{Jackiw:2003pm}
\begin{equation}\label{j10}
   Z_{\pm} = \int_{\Sigma_\rho}\, d^3y\, \epsilon^{abc} \left(\frac12\, \Gamma^i_{aj}\partial_b \Gamma^j_{ci}+\frac13\, \Gamma^i_{aj}\Gamma^j_{bk}\Gamma^k_{ci}\right)
\end{equation}
This expression can be checked to be diffeomorphism invariant,
because its variation with respect to $\gamma$ produces the Cotton
tensor, which is symmetric, traceless, with vanishing covariant
divergence.

We can now easily check that this solution is self-consistent.
Indeed letting the first term in (\ref{j7}) act on $e^{\pm i S^1}\,
Z_{\pm}$ we obtain terms of order of $\rho^3$ and $\rho^2$ which are
small in the limit $\rho\rightarrow0$.

\section*{D=4}

For $D=4$ the correspondence is AdS$_5$/CFT$_4$. The Wheeler-De Witt equations is now given by
\begin{equation}\label{j11}
H_\rho =-\kappa^2 \rho^{8}\,\hP\circ_{\K}\hP(x) + \rho^2\, R[\gamma](x) +
\frac{12}{\ell^2}\,,
\end{equation}
and the wave function can still be written as in Eq.(\ref{j4}),
with
\begin{equation}\label{j12}
    S^0=\frac{3}{\kappa\ell\rho^4} \int_{\Sigma_\rho} \sqrt{\gamma}\, d^3x\,.
\end{equation}
The analogous of Eq.(\ref{j7}) now reads
\begin{equation}\label{j13}
\bigg[\kappa^2 \rho^{6}\hP\circ_{\K}\hP(x) -  R[\gamma](x)\mp
2\frac{i\kappa\rho^2}{\ell}\, \hP(x) \bigg] e^{\pm i S^1}\, Z_\pm =0
\,.
\end{equation}
In order to cancel the $R$-term, let us make the following ansatz
\begin{equation}\label{j14}
   S^1 =\frac{\ell}{4\kappa\rho^2}\, \int_{\Sigma_\rho} \sqrt{\gamma}\, R[\gamma]
   d^3y\,.
\end{equation}
The first term in Eq.(\ref{j13}) cannot be neglected anymore and the resulting equation for $Z_\pm$ now reads
\begin{equation}\label{j15}
\frac{\ell^2}{4}\,\left(R^a_b(x)R^b_a(x) - \frac{1}{3}R^2(x)\right)Z_{\pm}\mp
2\frac{i\kappa}{\ell}\,\hP\, Z_{\pm} =0\,,
\end{equation}
which coincides with the conformal Ward identity in $D=4$ for proper values of the central charges. An explicit solution for the equation above has been given in \cite{Freidel:2008sh} and it takes the same form as $Z_\pm$ in Eq.(\ref{25}) with the Liouville action given by
\begin{equation}
S_{L}(\phi,\gamma)=\int_{\Sigma_\rho} d^3x\left(\frac12 \phi\nabla^2_4\phi+\phi Q_4(\gamma)\right),
\end{equation}
in which
\begin{equation}
\nabla^2_4=\nabla^2+\nabla_a(2P\gamma^{ab}-4P^{ab})\nabla_b,\qquad Q_4(\gamma)=\frac{1}{2}\nabla^2P+P^2-P^a_bP^b_a,
\end{equation}
with $P_{ab}=\frac{1}{2}\left(R_{ab}-\frac{1}{3}\gamma_{ab}R\right)$.

\section*{Loop corrections}

In what follows, we are going to investigate the asymptotic behavior
of those terms which are absent in the  un-regularized scheme
presented in \cite{Freidel:2008sh} and which are peculiar of the
adopted regularization procedure. Since they are purely quantum
corrections just like loop corrections in ordinary quantum field
theory, we will call them `loop corrections'.

For a generic dimension $D>2$ the expansion given in \cite{Freidel:2008sh} is consistent. In fact, the zero order term reads
\begin{equation}\label{13}
  \Psi^{(0)}_\rho(\gamma)=\exp{\left(\pm\frac{i}{\kappa \rho^D}\frac{D-1}{\ell}\int_{\Sigma_\rho} \sqrt{\gamma(x)}d^Dx\right)},
\end{equation}
such that
\begin{equation}\label{16}
\kappa^2 \rho^{2D}\,\hP\circ_{\K}\hP(x)\, \Psi^{(0)}_\rho=\frac{D(D-1)}{\ell^2}\Psi^{(0)}_\rho.
\end{equation}
The first order expansion of Eq.(\ref{10}) reads
\begin{equation}\label{18}
\pm i\frac{2\kappa}{\ell}\rho^D\hP(x)\Psi^{(1)}_\rho+\rho^2R[\gamma]\Psi^{(1)}_\rho=\kappa^2 \rho^{2D}\,\hP\circ_{\K}\hP(x)\,\Psi^{(1)}_\rho,
\end{equation}
and since the right-hand side can be neglected in the limit $\rho\rightarrow0$, a proper solution is given by
\begin{equation}\label{17}
  \Psi^{(1)}_\rho(\gamma)=\exp{\left(-i\frac{1}{\kappa}\rho^{2-D}\frac{\ell}{2(D-2)}\int_{\Sigma_\rho} \sqrt{\gamma(x)}R[\gamma]d^Dx\right)}.
\end{equation}

The right-hand side of Eq.(18) enters next orders of the asymptotic expansion and it gives
$$\kappa^2
\rho^{2D}\,\hP\circ_{\K}\hP(x)\,\Psi^{(1)}_\rho=\frac{\ell^2}{(D-2)^2}\rho^{4}\,\left(R^a_b(x)R^b_a(x)
- \frac{D}{4(D-1)}R^2(x)\right)\Psi^{(1)}_\rho-$$
$$-\frac{i\kappa\ell}{(D-2)}\rho^{2+D}\frac{D^2-2}{D-1}\K(x,x;t)R(x)\Psi^{(1)}_\rho
+$$
\begin{equation}
+\frac{i\kappa\ell}{D-2}(2-D(D-1))\rho^{2+D}[\nabla^2_{(y)}K(x,y,t)]|_{y=x}\Psi^{(1)}_\rho.\label{20}
\end{equation}
The first term coincides with the one found in
\cite{Freidel:2008sh}, while the additional contributions are
corrections due to renormalization scheme. In fact, in the
coincidence limits $\K(x,x;t)$ and
$[\nabla^2_{(y)}\K(x,y;t)]|_{y=x}$ the singularities  encountered as
$t\rightarrow0$ give rise to the finite renormalization constants
\cite{Mansfield:1994}, \cite{KowalskiGlikman:1996ad}.

At the next order, the Wheeler--De Witt equation reads
$$\pm i\frac{2\kappa}{\ell}\rho^D\hP(x)\Psi^{(2)}_\rho=(\Psi^{(1)}_\rho)^{-1}\left[\kappa^2
\rho^{2D}\,\hP\circ_{\K}\hP(x)\,\Psi^{1}_\rho\right]\Psi^{(2)}_\rho=$$
\begin{equation}
=\frac{\ell^2}{(D-2)^2}\rho^{4}\,\left(R^a_b(x)R^b_a(x) -
\frac{D}{4(D-1)}R^2(x)\right)\Psi^{(2)}_\rho+O(\rho^5),
\end{equation}
and the loop corrections are negligible being $O(\rho^{2+D})$. This is the case for all the relevant orders as $\rho\rightarrow0$. In fact, for a generic term of order $n$ in the expansion one has the following equation in the un-regularized case
\begin{equation}
\pm i\frac{2\kappa}{l}\rho^D\hP(x)\Psi^{(n)}_\rho\propto \rho^{2n}\Psi^{(n)}_\rho+O(\rho^{2n+1}).
\end{equation}
The leading corrections coming from the regularization procedure are $O(\rho^{D+2})$ (\ref{20}), thus they are relevant for $2n\geq D+2$. At such orders of the asymptotic expansion one has
\begin{equation}
\hP(x)\Psi^{(n)}_\rho\propto \rho^{2n-D}\Psi^{(n)}_\rho,\qquad 2n-D\geq 2,
\end{equation}
and being the dependence from $\rho$ through a positive power, the right-hand side of the equation above can be neglected as $\rho\rightarrow0$. Hence, the terms of the expansion for which loop corrections are relevant turn out to be negligible in the asymptotic limit.

\section*{Conclusions}

In this work, we presented a regularization scheme for
the Wheeler-De Witt  operator describing the radial evolution of the
gravitational field in $D+1$-dimensional spacetime. This way, we
could investigate the role of loop corrections on the asymptotic
behavior of the wave function in an asymptotically AdS space time.
This analysis was motivated by the results presented in
\cite{Freidel:2008sh}, which elucidated the correspondence between
the bulk wave function and the partition function of a boundary CFT
via the asymptotic expansion of the solution of the radial
Wheeler-De Witt equation. We found that all the corrections coming
from renormalization are negligible in the asymptotic limit.
Therefore, the asymptotic form of the wave-function $\Psi$ is
insensitive to loop corrections and the results of
\cite{Freidel:2008sh} are well-grounded. Although this result is not surprising in view of \cite{Papadimitriou:2010as},
where it was argued that the
solution of the WKB expansion gets asymptotical contributions
only from the long-distance divergencies of the gravitational action, it is encouraging to derive it explicitly.

In this paper we considered only pure (quantum) gravity in the bulk and we showed
that the asymptotic wave function is conformally invariant (up to the possible conformal anomaly).
This result holds for an arbitrary quantum gravity theory described, at least in some regime, by Wheeler-De Witt equation,
 irrespectively
of its possible UV completion. If so the theory should somehow
notice the renormalizability problem of pure quantum gravity. This
problem should disappear, of course, in the UV complete theory, like
string theory, in which the short distance pathologies are
automatically taken care of. However, as explained above, even in
pure quantum gravity the renormalizability problem arises only when
we move away from the boundary, and the terms of higher order in
$\rho$ cannot be neglected anymore. Since the radial flow in the
bulk is associated with the renormalization group flow of the dual
theory \cite{deBoer:2000cz}, \cite{de Boer:1999xf}, \cite{rgflow}
(see \cite{rgflow2} for recent developments), the running of the
couplings of the conformal theory should be sensitive to loop
corrections in the bulk. The connection between the radial flow of
the bulk wave function and the renormalization group flow of the
theory on the boundary will be the subject of our future
investigations.

It should be mentioned that Wheeler-De Witt equation has already been used in \cite{Kubota:2003pf}
to evaluate the $1/N^2$ corrections to the conformal anomaly of the boundary
$\mathcal{N}=4$ $SU(N)$ super-Yang-Mills theory in
AdS$_5$/CFT$_4$. The results coincided with the ones obtained in
\cite{Mansfield:2002pa} in the Schrodinger representation. It is worth noting
the crucial role of the heat-kernel expansion in such calculations
(the anomaly corrections are proportional to the coefficient $a_2(x,x)$ in the
De Witt-Schwinger proper time representation). However,  \cite{Kubota:2003pf} discusses
the case of gravity coupled to $N$ massive scalar fields and the authors use the
heat kernel to regularize the kinetic part of the scalar field only, which makes it hard
to directly compare their results with ours.

\section*{Acknowledgments}
 JKG is supported in part by the grant
2011/01/B/ST2/03354; FC and JKG are supported by funds provided by
the National Science Center under the agreement
DEC-2011/02/A/ST2/00294.

\end{document}